\documentclass[sigconf]{acmart}

\usepackage{graphicx}
\usepackage{placeins}

\setcopyright{none}
\renewcommand\footnotetextcopyrightpermission[1]{}
\acmConference[RecSys '26]{The 20th ACM Conference on Recommender Systems}{September 28--October 2, 2026}{Minneapolis, MN, USA}
\acmYear{2026}
\acmBooktitle{Proceedings of the 20th ACM Conference on Recommender Systems (RecSys '26), September 28--October 2, 2026, Minneapolis, MN, USA}

\title{Ocean4Rec: Offline LLM-Derived OCEAN Profiles for Request-Time VOD Reranking}

\author{Wonkyun Kim}
\affiliation{%
  \institution{Samsung Electronics}
  \country{Republic of Korea}
}
\email{wkyun.kim@samsung.com}

\author{Sehyun Bae}
\affiliation{%
  \institution{Samsung Electronics}
  \country{Republic of Korea}
}
\email{sehyun.bae@samsung.com}

\author{Kwanki Ahn}
\affiliation{%
  \institution{Samsung Electronics}
  \country{Republic of Korea}
}
\email{kwanki.ahn@samsung.com}

\author{Mungyu Bae}
\affiliation{%
  \institution{Samsung Electronics}
  \country{Republic of Korea}
}
\email{mung.bae@samsung.com}

\author{Saeun Choi}
\affiliation{%
  \institution{Samsung Electronics}
  \country{Republic of Korea}
}
\email{saeun.choi@samsung.com}

\author{Soyeon You}
\affiliation{%
  \institution{Samsung Electronics}
  \country{Republic of Korea}
}
\email{soyeon.you@samsung.com}

\author{Chandra Prabhakar}
\affiliation{%
  \institution{Samsung Electronics}
  \country{India}
}
\email{chandra.pr@samsung.com}

\author{Sehyun Kim}
\authornote{Corresponding author.}
\affiliation{%
  \institution{Samsung Electronics}
  \country{Republic of Korea}
}
\email{sehyun216.kim@samsung.com}

\ccsdesc[500]{Information systems~Recommender systems}
\ccsdesc[100]{Computing methodologies~Neural networks}
\keywords{recommender systems, reranking, large language models, OCEAN, video-on-demand}

\begin{document}

\begin{abstract}
Industrial video-on-demand (VOD) recommenders need richer content understanding, but LLM-as-reranker designs repeat prompt construction, token generation, model invocation, output parsing, and fallback handling for each request.
In high-volume latency-sensitive services, these request-time operations complicate throughput planning, tail-latency control, capacity isolation, and predictable operation.
This paper presents Ocean4Rec, a reranking layer that uses an LLM only offline to materialize item OCEAN profiles from content metadata.
Items are mapped into Openness, Conscientiousness, Extraversion, Agreeableness, and Neuroticism scores, while user profiles are built by time-decayed aggregation of recently clicked and deep-linked items in the same five-dimensional space.
At request time, Ocean4Rec joins precomputed item profiles, user profiles, base recommender scores, and catalog recency, then performs numeric reranking without an LLM call.
On anonymized Samsung Smart TV VOD logs, same-candidate Top1000 temporal-holdout offline evaluations show that Ocean4Rec improves NDCG@20 over a stronger non-OCEAN Base+Recency ordering by 7.6\% for an NCF generator and 61.5\% for a LightGCN generator.
HR@20 is inconclusive for NCF and improves by 67.3\% for LightGCN, reflecting sparse exact-item replay labels and the strength of recency as an industrial baseline.
The result should be read as offline replay evidence for a bounded auxiliary content-taste feature that preserves the deployability advantage of a request-time-LLM-free serving path.
\end{abstract}

\maketitle

\section{Introduction}

Production recommenders are served through multi-stage pipelines: candidate generation, ranking or reranking, filtering, and final list construction \cite{cheng2016wide,covington2016deep,gomezuribe2015netflix}.
These systems must satisfy constraints that are often secondary in research-track evaluations: predictable latency, feature freshness, model availability, privacy boundaries, incident recovery, and operator-facing diagnosis.
LLMs can enrich recommendation through feature extraction, user modeling, item representation, explanation, and candidate scoring \cite{zhao2024recommender,lin2024benefit,wu2024survey}.
Recent LLM rankers for recommendation and retrieval ask generative models to score or permute candidates through zero-shot prompting, instruction tuning, pairwise prompting, or listwise reranking \cite{hou2024llmrank,sun2023rankgpt,ma2023listwise,qin2024prp,pradeep2023rankvicuna,luo2024recranker}.
Other LLM-based recommenders, including TALLRec and EXP3RT, align LLMs with recommendation data or use review-driven preference reasoning to improve prediction and reranking \cite{bao2023tallrec,kim2025exp3rt}.
Persona4Rec is especially relevant because it shifts item-side reasoning offline and reports competitive accuracy against recent LLM-based rerankers while reducing inference time \cite{kim2026persona4rec}.
Together, these studies show that LLMs can provide useful recommendation judgments, but direct request-time use moves candidate serialization, prompt construction, token generation, model invocation, output parsing, and retry logic into the serving path.
Systems work on LLM serving shows that high-throughput generation needs careful batching, scheduling, and memory management \cite{yu2022orca,kwon2023pagedattention}.
In a high-traffic VOD surface, the operational burden is not only average latency; it also includes tail latency, GPU or API capacity planning, queueing behavior, rate-limit handling, incident blast radius, and fallback design.

Ocean4Rec studies a different placement.
An LLM is used as an offline annotator for VOD metadata, not as a request-time reasoning engine.
The generated item profiles are stored as numeric OCEAN features.
User OCEAN taste profiles are then computed from recent implicit interactions with profiled content.
At request time, the system retrieves scored candidates from an existing recommender, joins precomputed OCEAN and recency features, computes a low-dimensional score, and returns a reranked list.
No prompt is built and no LLM is invoked in the request path.

The design is intentionally conservative.
Implicit-feedback methods such as matrix factorization, neural collaborative filtering, and graph collaborative filtering remain strong interaction-driven relevance estimators \cite{hu2008implicit,rendle2009bpr,he2017neural,he2020lightgcn}.
Content-based and hybrid recommender work has long combined item attributes with collaborative evidence \cite{burke2002hybrid,adomavicius2005toward,lops2011content}.
Ocean4Rec follows that late-stage hybrid pattern, but uses an LLM-derived, schema-constrained OCEAN content space that is shared by users and items without learned alignment.
Compared with direct LLM rankers, it removes request-time LLM inference and keeps the online path bounded by ordinary feature lookup, low-dimensional scoring, and sorting.
Compared with dense metadata embeddings or free-form personas, it gives operators a five-dimensional, inspectable feature that can fail open to the base score.
Compared with recency-only boosting, it adds a content-compatibility signal tied to recent user behavior.

\begin{figure*}[t]
\centering
\includegraphics[width=\textwidth]{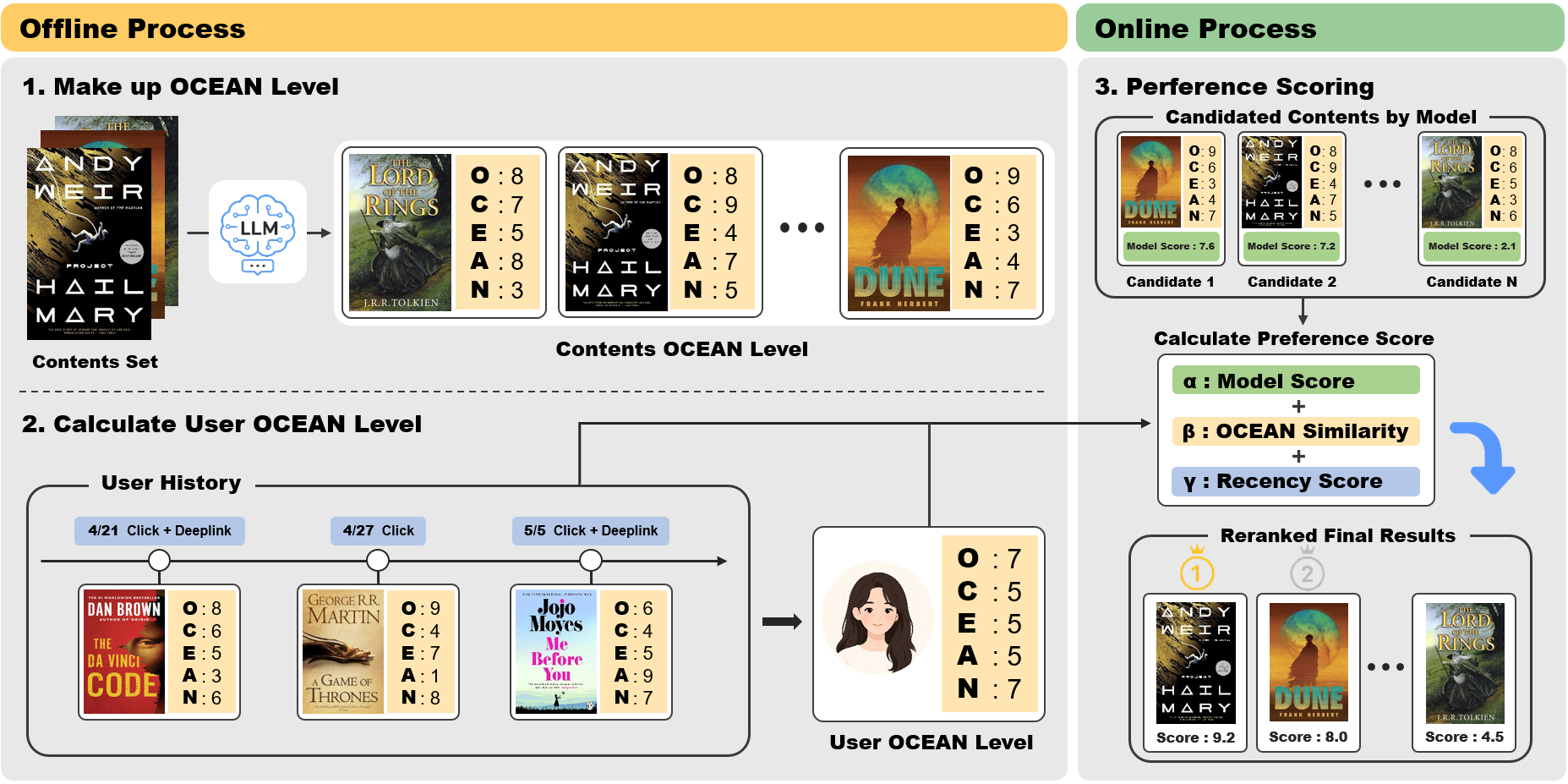}
\caption{Ocean4Rec overview. Item OCEAN profiles are generated offline from content metadata, user OCEAN profiles are computed from recent profiled interactions, and request-time reranking combines base model score, OCEAN compatibility, and catalog recency without an online LLM call.}
\Description{An overview of the Ocean4Rec pipeline. The offline process creates item OCEAN profiles with an LLM and aggregates user history into a user OCEAN profile. The online process scores candidate content by combining model score, OCEAN similarity, and recency score, then returns reranked results.}
\label{fig:ocean4rec-overview}
\end{figure*}

The paper makes three contributions for the RecSys Industry Track.
First, it describes an offline LLM materialization pattern for adding OCEAN content profiles to an industrial VOD reranker without request-time LLM calls.
Second, it defines a same-space user/item representation in which items are profiled from metadata and users are represented by time-decayed aggregation of consumed item profiles.
Third, it reports same-candidate temporal-holdout offline evaluations on anonymized Samsung Smart TV VOD logs with NCF and LightGCN candidate generators.
All empirical claims are scoped to offline point estimates under the evaluated generators and fixed score settings.
Figure \ref{fig:ocean4rec-overview} summarizes the offline materialization and request-time reranking flow.

\section{Industrial Design Goals}

\textbf{Request-time footprint and stability.}
The recommender surface cannot depend on per-request LLM inference, token generation, or prompt-size-sensitive runtime.
The online reranker must use ordinary feature lookup, numeric scoring, and sorting so that throughput and fallback behavior are predictable.

\textbf{Base-model compatibility.}
The reranker consumes scored candidate lists from existing generators.
It does not require retraining or modifying every base recommender before experimentation.

\textbf{Missing-profile tolerance.}
Catalog metadata quality is uneven.
Items with missing or invalid OCEAN profiles remain eligible through the base score rather than becoming unavailable.

\textbf{Operator traceability.}
Operators should be able to inspect why an item moved upward: recent profiled history, the user OCEAN vector, the item OCEAN vector, the similarity term, and the recency term.
This is an operator-facing diagnostic, not a claim of psychological assessment.

\textbf{Privacy boundary.}
The LLM-facing batch step receives item-side metadata only.
No user identifier, account identifier, device identifier, user history, exposure count, popularity count, or evaluation-window label is sent to the LLM.
Behavior-derived user OCEAN vectors are internal recommendation features governed as recommender-system data, not user-facing personality labels.

\section{Ocean4Rec Method}

\subsection{Offline Item OCEAN Profiles}

Ocean4Rec annotates each eligible VOD item once with an LLM.
The evaluated profile-generation run used Google Gemini 2.5 Pro as the offline annotator \cite{google2025gemini25pro}.
The prompt maps title, plot, external plot when available, description, and optional genre metadata to a schema-constrained JSON record containing the item identifier, five integer OCEAN scores, confidence, and a short reason.
The five dimensions follow the Big Five taxonomy: Openness, Conscientiousness, Extraversion, Agreeableness, and Neuroticism \cite{goldberg1990bigfive,costa1992neo,john1999bigfive}.
Personality-aware recommenders have studied explicit personality traits as user signals \cite{hu2011personality,tkalcic2015personality}; Ocean4Rec makes a narrower use of the vocabulary as an item-content rubric.
The scores describe whether metadata emphasizes novelty and imagination, structure and goal orientation, social energy, interpersonal warmth, or emotional intensity.

The offline pipeline validates generated records before ingestion.
Malformed outputs, missing fields, score-range violations, and identifier mismatches trigger retry or chunk splitting.
Unrecoverable single-item failures receive low-confidence neutral fallback profiles.
Items with no eligible textual metadata remain in the base candidate set, and the request-time reranker omits the OCEAN term for those candidates.

\subsection{User OCEAN Taste Profiles}

User profiles are constructed from recent interactions with profiled items.
For user \(u\), let \(H_u\) be the set of historical content-click and deep-link/select-source interactions before the profile cutoff.
Each interaction links the user to an item OCEAN vector \(q_i\), timestamp \(t\), and event type \(e\).
Recent interactions receive exponential half-life decay:
\[
  w_{\mathrm{time}}(a)=0.5^{a/h},
\]
where \(a\) is interaction age in days and \(h\) is the half-life.
The reported implementation uses a 90-day lookback window and 90-day half-life; with the March 31, 2026 cutoff, the profile window starts on January 1, 2026.
The use of time decay follows the long-standing observation that recommender signals can change with temporal dynamics \cite{koren2009temporal}.
Watch-duration and completion signals are not used in the reported profile.

The user profile is the weighted average:
\[
  z_u =
  \frac{\sum_{(i,t,e)\in H_u}w_{\mathrm{time}}(a_t)q_i}
       {\sum_{(i,t,e)\in H_u}w_{\mathrm{time}}(a_t)}.
\]
The reported implementation uses event type for filtering, traceability, and monitoring, but it does not assign different profile weights to clicks and deep-link/select-source events.
Interactions whose items do not have valid OCEAN profiles are skipped in profile construction.
Repeated interactions with the same item are collapsed at the user-item level by keeping the highest-weight and most recent row.
This prevents one frequently clicked title from dominating the user vector through duplicate events.

\subsection{Reranking Score}

For candidate item \(i\), Ocean4Rec combines base relevance, OCEAN compatibility, and catalog recency:
\[
  S(u,i)=\alpha B(u,i)+\beta P(u,i)+\gamma R(i).
\]
Here \(B(u,i)\) is the base recommender score min--max normalized within the user's candidate set.
If candidate scores are unavailable or degenerate, the implementation falls back to a rank-derived base feature.
The OCEAN term is Pearson correlation between \(z_u\) and \(q_i\), clipped and transformed from \([-1,1]\) to \([0,1]\):
\[
  P(u,i)=\frac{\mathrm{clip}(\rho(z_u,q_i),-1,1)+1}{2}.
\]
Pearson-style correlation has long been used to compare preference patterns in collaborative filtering \cite{resnick1994grouplens}.
Because OCEAN has only five dimensions, low-variance vectors require explicit handling: the implementation assigns correlation \(1.0\) only when the two vectors are identical and \(0.0\) otherwise.
Non-degenerate correlations are clipped before the score transformation.

The recency term is:
\[
  R(i)=0.5^{\mathrm{ContentReleaseAgeDays}_i/365}.
\]
Content release age is computed relative to the March 31, 2026 cutoff; negative ages are clipped to zero and missing release dates receive recency score zero.
The reported experiments use fixed internal convex score settings chosen from pre-evaluation internal configuration artifacts and frozen before the April evaluation window was scored.
Evaluation-window labels are not used for coefficient selection.
For component rows, the omitted auxiliary term receives zero weight and the remaining nonzero weights are renormalized within the same internal configuration family.
The non-sensitive constraints are that the base-score term is the largest single component, the OCEAN term is a capped auxiliary component, and all active components use nonnegative normalized weights.
Exact coefficients and the internal selection artifact are not released because they are tied to business-sensitive serving configuration.
Accordingly, the rows are fixed-setting diagnostics under identical candidate sets, labels, and temporal split, not a leakage-auditable public tuning recipe.

If a user or item lacks a usable OCEAN profile, the OCEAN weight is reassigned to the base score:
\[
  S(u,i)=(\alpha+\beta)B(u,i)+\gamma R(i).
\]
This fallback reduces the auxiliary signal but does not remove candidates.

\subsection{Request-Time Architecture}

The request-time path is base-candidate retrieval, feature lookup, \(O(|C_u|d)\) OCEAN scoring with \(d=5\), weighted summation, sorting, and Top-\(k\) selection.
The operational dependency graph is therefore: offline LLM item-profile generation, batch validation and storage, user-profile refresh from historical events, online candidate retrieval, feature join, reranking, and fallback to base-score ordering when a profile is missing.
Compared with online LLM reranking, this removes prompt construction, token generation, model execution, output parsing, and LLM-specific retry handling from serving.
Total cost accounting depends on catalog size, batch cadence, provider, model choice, and serving stack.
We therefore do not report measured latency, infrastructure cost, API price, or online availability.
Any cost-related claim in this paper is limited to the architectural property that the evaluated reranking path has zero request-time LLM calls.

\begin{table}[t]
\centering
\scriptsize
\setlength{\tabcolsep}{3pt}
\begin{tabular}{@{}p{0.35\columnwidth}p{0.57\columnwidth}@{}}
\toprule
Request-time step & Evaluated path \\
\midrule
LLM call & None; LLM use is offline item profiling only. \\
Candidate input & Existing scored Top1000 list from a base generator. \\
Feature payload & One user OCEAN vector, item OCEAN vector, base score, and release-age recency per candidate. \\
Scoring work & CPU-compatible five-dimensional similarity, scalar blending, and sorting. \\
Fallback & Missing OCEAN feature reassigns the OCEAN weight to the base score; base ordering remains available. \\
Deployment control & OCEAN contribution can be capped or disabled independently of the base generator. \\
\bottomrule
\end{tabular}
\caption{Non-sensitive request-time footprint. The table reports architectural work, not measured latency or cost.}
\label{tab:deployment-footprint}
\end{table}

\section{Evaluation Setup}

\subsection{Commercial VOD Data}

We evaluate Ocean4Rec with anonymized interaction logs from a commercial connected-TV VOD environment.
Under privacy and business constraints, we disclose only that the data originates from a Samsung Smart TV-based VOD setting.
We do not disclose internal service names, raw log table names, account identifiers, device identifiers, exact traffic volume, or row-level catalog snapshot identifiers.
As non-sensitive scale bands, the evaluated slice contains more than 80k candidate-bearing user identifiers per generator, more than 40k evaluated users per generator after future-label filtering, more than 10k candidate items per generator, hundreds of thousands of future-active user identifiers, and more than one million future click/deep-link events.

The logs contain implicit feedback generated in a lean-back TV context, including content clicks, detail-page visits, deep-link visits, and VOD watch events.
The reported relevance labels are future content-click and deep-link events.
The content catalog provides title, plot, description, genre, and release-date metadata used for item understanding, eligibility filtering, and recency.
The OCEAN-profile artifact covers the eligible VOD catalog used by the evaluated candidate generators and includes high-, medium-, and low-confidence profiles as well as neutral fallback profiles.
Low-confidence and neutral fallback profiles are included with the same OCEAN weight in the reported configuration; confidence-aware down-weighting is left to future work.

\subsection{Temporal Holdout and Leakage Boundary}

We use a global temporal holdout to mimic a production recommendation scenario.
User profiles and candidate lists are generated using interactions and metadata available up to March 31, 2026.
Recommendation quality is evaluated against future interaction events observed from April 1, 2026 to April 27, 2026.
This protocol differs from common leave-one-out protocols; it is selected because time-overlapping splits can introduce temporal leakage in deployment-oriented settings \cite{ji2023leakage,gusak2025split}.

The temporal boundary is enforced at the artifact level.
User OCEAN profiles use content-click and deep-link/select-source events from January 1, 2026 through March 31, 2026 only.
NCF and LightGCN Top1000 candidate lists and base scores are generated for the March 31 cutoff.
The recency feature converts catalog release date to age relative to March 31 and does not use post-cutoff interaction, exposure, popularity, or editorial-update signals.
LLM item-profile prompts contain no user events, exposure counts, popularity counts, or evaluation-window labels.
Evaluation labels are consumed only by the evaluation job.
We do not publish raw catalog snapshot identifiers or metadata update timestamps, so the audit excludes post-cutoff behavioral and popularity signals rather than providing a full raw-catalog replay.

\subsection{Candidate Generators}

The completed base recommender experiments use NCF and LightGCN candidate generators.
Ocean4Rec is not tied to either model because it consumes scored candidate lists and precomputed item/user features.
The NCF generator is a production-style GMF/MLP NCF model trained on implicit VOD rating rows available at the March 31 cutoff.
Its candidate export scores eligible items, keeps a raw Top2000 list, removes recently watched titles and a small fixed exclusion list for the evaluated slice, and writes final Top1000 scored candidates.
The LightGCN generator is trained for the same cutoff from binary positive implicit VOD interactions with 64-dimensional embeddings, three propagation layers, one sampled negative per positive, and early stopping against validation HitRate@10.
Both candidate generators are frozen before Ocean4Rec reranking.

\subsection{Baselines and Metrics}

The same-candidate protocol compares four orderings over each generator's Top1000 candidate set: Base, Base+Recency, Base+OCEAN, and Ocean4Rec.
This isolates reranking effects from candidate-width expansion.
All score settings are fixed before April labels are consumed; future labels are used only after ranking for metric computation.
Metrics are computed only for future-active users with both candidates and future-window labels.
These values are therefore not production CTR, all-traffic utility, or exposure-conditioned utility.

We report standard offline ranking metrics \cite{jarvelin2002cumulated,herlocker2004evaluating,shani2011evaluating}: HR@\(k\), MRR@\(k\), and NDCG@\(k\) for \(k\in\{10,20\}\).
Metrics are computed per evaluated user and then averaged.
HR@\(k\) is the fraction of evaluated users with at least one exact item-level hit in the Top-\(k\) list.
MRR@\(k\) is the reciprocal rank of the first hit within Top-\(k\), or zero when no hit exists.
Future labels are de-duplicated to unique content IDs per user.
NDCG@\(k\) uses binary relevance with ideal DCG computed from \(\min(k,\) the user's unique future-label count\()\).
Because relevance requires exact content-ID matches to future click or deep-link events, the absolute metric scale is low; we therefore interpret the results as sparse offline replay point estimates rather than CTR-like business rates.
For highlighted deltas, we report 95\% paired bootstrap confidence intervals from 500 resamples over evaluated users \cite{efron1993bootstrap}.
These intervals describe user-level offline replay variation under the fixed log slice; they do not imply online product impact.

\section{Results}

\subsection{Same-Candidate Ablation}

Table \ref{tab:main-ablation} reports the same-candidate ablation.
For NCF, Ocean4Rec has the highest HR and NDCG point estimates at @10 and @20, while Base has the highest MRR.
For LightGCN, Ocean4Rec has the highest point estimates for all reported metrics at both cutoffs.
Relative to the stronger Base+Recency row, Ocean4Rec improves NDCG@20 by 7.6\% for NCF and 61.5\% for LightGCN.
The HR@20 movement is small for NCF and larger for LightGCN.
Relative to base-score-only sorting, Ocean4Rec has higher HR@20 and NDCG@20 by 21.8\% and 15.0\% for NCF, and by 74.6\% and 67.9\% for LightGCN.
The absolute @20 deltas versus Base are +0.0092 HR and +0.0009 NDCG for NCF, and +0.0194 HR and +0.0022 NDCG for LightGCN.
Against Base+Recency, the relative 95\% bootstrap CIs for Ocean4Rec's NDCG deltas are +11.3\% to +23.0\% at @10 and +3.6\% to +12.1\% at @20 for NCF, and +31.7\% to +66.7\% at @10 and +45.8\% to +77.6\% at @20 for LightGCN.
For HR@20, the corresponding relative CIs are -1.9\% to +4.0\% for NCF and +55.7\% to +78.8\% for LightGCN; for MRR@20, they are +5.6\% to +15.7\% for NCF and +24.7\% to +54.4\% for LightGCN.

\begin{table*}[t]
\centering
\scriptsize
\setlength{\tabcolsep}{3pt}
\begin{tabular*}{\textwidth}{@{\extracolsep{\fill}}lllrrr@{}}
\toprule
Gen. & \(k\) & Order & HR & MRR & NDCG \\
\midrule
NCF & 10 & Base & 0.0263 & \textbf{0.0120} & 0.0050 \\
NCF & 10 & Base+Rec. & 0.0289 & 0.0102 & 0.0048 \\
NCF & 10 & Base+OCEAN & 0.0281 & 0.0095 & 0.0046 \\
NCF & 10 & Ocean4Rec & \textbf{0.0333} & 0.0116 & \textbf{0.0055} \\
NCF & 20 & Base & 0.0422 & \textbf{0.0131} & 0.0057 \\
NCF & 20 & Base+Rec. & 0.0510 & 0.0116 & 0.0061 \\
NCF & 20 & Base+OCEAN & 0.0450 & 0.0106 & 0.0055 \\
NCF & 20 & Ocean4Rec & \textbf{0.0514} & 0.0129 & \textbf{0.0065} \\
LightGCN & 10 & Base & 0.0162 & 0.0061 & 0.0028 \\
LightGCN & 10 & Base+Rec. & 0.0166 & 0.0062 & 0.0028 \\
LightGCN & 10 & Base+OCEAN & 0.0167 & 0.0063 & 0.0029 \\
LightGCN & 10 & Ocean4Rec & \textbf{0.0277} & \textbf{0.0084} & \textbf{0.0042} \\
LightGCN & 20 & Base & 0.0259 & 0.0067 & 0.0032 \\
LightGCN & 20 & Base+Rec. & 0.0271 & 0.0069 & 0.0033 \\
LightGCN & 20 & Base+OCEAN & 0.0263 & 0.0069 & 0.0032 \\
LightGCN & 20 & Ocean4Rec & \textbf{0.0453} & \textbf{0.0097} & \textbf{0.0054} \\
\bottomrule
\end{tabular*}
\caption{Same-candidate Top1000 temporal-holdout ablation. Values are rounded offline point estimates within the same generator and cutoff; relative deltas in text use full precision.}
\label{tab:main-ablation}
\end{table*}

The component rows define the practical interpretation.
Base+Recency has higher NDCG@20 point estimates than Base for both generators in this slice, which is consistent with catalog recency acting as a temporal demand prior \cite{koren2009temporal}.
Base+OCEAN is generator-dependent: it raises HR@20 for both generators but lowers NCF MRR@20 and NDCG@20.
Ocean4Rec is strongest when OCEAN is paired with base relevance and recency.
This supports OCEAN as a bounded auxiliary compatibility feature, not a standalone ranker.

\subsection{Secondary Diagnostics}

Table \ref{tab:secondary} reports secondary diagnostics.
We include a Persona4Rec diagnostic because Persona4Rec reports competitive LLM-based recommendation accuracy while preserving a request-time-LLM-free scoring path \cite{kim2026persona4rec}.
For the VOD diagnostic, our metadata-only Persona4Rec implementation uses Google Gemini 2.5 Pro to generate item personas from VOD metadata offline \cite{google2025gemini25pro}; embeds personas with a frozen BGE-M3 encoder \cite{chen2024m3embedding}; builds user vectors as time-decayed averages of recent item-persona embeddings; and sorts candidates by semantic similarity without base-score or recency combination.
This differs from the original review-grounded trained Persona4Rec setup because the VOD catalog does not provide public item reviews in the evaluated serving path; the diagnostic tests the offline persona-indexing idea under connected-TV metadata constraints.
Because this metadata-only Persona4Rec diagnostic does not include the same base-score and recency fusion used by Ocean4Rec, it should not be read as a fully matched head-to-head baseline.
The public diagnostics use sampled Amazon-Books and Yelp subsets \cite{hou2024bridging,amazonReviews2023,yelpOpenDataset}.
They are included as sampled protocol sanity checks outside the connected-TV VOD slice; they do not support public-benchmark generality claims.

\begin{table*}[t]
\centering
\scriptsize
\setlength{\tabcolsep}{2pt}
\begin{tabular*}{\textwidth}{@{\extracolsep{\fill}}llrrrrrr@{}}
\toprule
Setting / Method & Gen. & HR@10 & MRR@10 & NDCG@10 & HR@20 & MRR@20 & NDCG@20 \\
\midrule
VOD / Base & NCF & 0.0263 & \textbf{0.0120} & 0.0050 & 0.0422 & \textbf{0.0131} & 0.0057 \\
VOD / Persona4Rec diag. & NCF & 0.0181 & 0.0065 & 0.0032 & 0.0312 & 0.0073 & 0.0039 \\
VOD / Ocean4Rec & NCF & \textbf{0.0333} & 0.0116 & \textbf{0.0055} & \textbf{0.0514} & 0.0129 & \textbf{0.0065} \\
\midrule
VOD / Base & LightGCN & 0.0162 & 0.0061 & 0.0028 & 0.0259 & 0.0067 & 0.0032 \\
VOD / Persona4Rec diag. & LightGCN & 0.0169 & 0.0051 & 0.0028 & 0.0290 & 0.0060 & 0.0035 \\
VOD / Ocean4Rec & LightGCN & \textbf{0.0277} & \textbf{0.0084} & \textbf{0.0042} & \textbf{0.0453} & \textbf{0.0097} & \textbf{0.0054} \\
\midrule
Amazon-Books / Base & BPR-MF & 0.0782 & 0.0265 & 0.0385 & 0.1385 & 0.0304 & 0.0534 \\
Amazon-Books / Ocean4Rec & BPR-MF & 0.0962 & \textbf{0.0412} & \textbf{0.0541} & 0.1385 & \textbf{0.0442} & \textbf{0.0649} \\
Amazon-Books / Persona4Rec diag. & BPR-MF & \textbf{0.1026} & 0.0388 & 0.0535 & 0.1385 & 0.0412 & 0.0624 \\
\midrule
Yelp / Base & LightGCN & 0.2460 & 0.0860 & 0.1233 & 0.3750 & 0.0950 & 0.1559 \\
Yelp / Ocean4Rec & LightGCN & \textbf{0.2540} & \textbf{0.0916} & \textbf{0.1292} & 0.3750 & \textbf{0.1000} & \textbf{0.1598} \\
Yelp / Persona4Rec diag. & LightGCN & 0.1830 & 0.0559 & 0.0850 & 0.3750 & 0.0687 & 0.1329 \\
\bottomrule
\end{tabular*}
\caption{Secondary diagnostics. VOD rows use the same temporal holdout and Top1000 candidate sets as Table \ref{tab:main-ablation}. Public rows use sampled public subsets and Top20 candidates, so absolute values must not be compared across VOD and public blocks. HR@20 on public Top20 rows is candidate-set recall rather than a rank-sensitive reranking metric. These rows are protocol sanity checks, not the main evidence.}
\label{tab:secondary}
\end{table*}

The VOD metadata-only Persona4Rec diagnostic does not match the base-score-aware Ocean4Rec setting on the temporal holdout.
In these sampled public diagnostic runs, Ocean4Rec has higher Amazon-Books MRR and NDCG than the BPR-MF base order, while Persona4Rec has higher HR@10 and all methods tie on HR@20.
On the sampled Yelp Restaurants diagnostic, Ocean4Rec has higher HR@10, MRR, and NDCG than the LightGCN base order.
In the public diagnostics, Amazon-Books shows positive relative bootstrap CIs for Ocean4Rec versus Base on MRR and NDCG: MRR@10 +14.7\% to +110.3\%, NDCG@10 +12.1\% to +77.3\%, MRR@20 +12.1\% to +87.0\%, and NDCG@20 +6.6\% to +37.4\%.
The sampled Yelp diagnostic has smaller ranking deltas, with relative CIs that include zero for NDCG@20 (-0.5\% to +5.4\%) and MRR@20 (-0.9\% to +11.8\%).
We treat these public results as sanity checks, not as the main evidence.

\section{Industry Lessons and Risk Controls}

\textbf{Move expensive LLM reasoning out of the serving loop.}
Ocean4Rec gets content understanding from an LLM while keeping the online path numeric and deterministic.
This is not a total-cost measurement, because offline batch frequency, model provider, catalog churn, and platform pricing all matter.
It does remove per-request LLM execution, prompt assembly, token generation, candidate-text serialization, output parsing, and LLM-specific failure handling from the evaluated recommender path.

\textbf{Keep collaborative relevance.}
The Base + OCEAN component is not uniformly better than Base.
The combined row keeps base relevance and recency, then adds a bounded OCEAN term.

\textbf{Use temporal evaluation for production-facing claims.}
The cutoff is designed to exclude April behavior, exposure, popularity, and labels from reranking features; raw catalog replay is not disclosed.
This matters in catalog-driven VOD settings where newly released or recently promoted titles can create temporal leakage.

\textbf{Govern OCEAN as a recommendation feature.}
The user vector is a behavior-derived aggregation of consumed item profiles.
It should not be surfaced as a psychological label, used as an identity attribute, or reused for advertising or sensitive targeting.
This boundary matters because digital behavioral traces can reveal private traits and support psychological targeting if governed incorrectly \cite{kosinski2013private,matz2017psychological}.
Before live launch, governance should include profile-retention controls, deletion handling under product policy, segment-level ranking audits, provider/category exposure checks, and operator review before increasing OCEAN weight.

\textbf{Audit fallback and metadata bias.}
OCEAN profiles reflect available metadata and the annotating model.
Metadata-sparse titles, genre stereotypes, language coverage, and recency-driven concentration should be monitored.
Operational monitors should track profile missing rate, neutral fallback rate, invalid-output retries, average OCEAN and recency score contribution, and provider/category exposure shifts.
Confidence-aware weighting and stricter fallback treatment remain deployment controls rather than measured results in this paper.

\begin{table}[t]
\centering
\scriptsize
\setlength{\tabcolsep}{3pt}
\begin{tabular}{@{}p{0.42\columnwidth}p{0.50\columnwidth}@{}}
\toprule
Monitor & Operational response \\
\midrule
Profile missing or neutral fallback spike & Reduce OCEAN contribution or fall back to base ordering until catalog audit. \\
Invalid LLM output retry increase & Hold profile publication and reuse the previous valid artifact or base order. \\
OCEAN contribution drift & Cap the auxiliary term or disable the reranker through configuration. \\
Provider or category exposure shift & Require operator review before increasing OCEAN weight. \\
\bottomrule
\end{tabular}
\caption{Deployment controls for an industrial launch. These are operational guardrails, not online impact measurements.}
\label{tab:deployment-controls}
\end{table}

\section{Limitations}

The evidence is offline.
We report future-window ranking metrics, but not online A/B impact, measured serving latency, infrastructure cost, API price, retention, satisfaction, or business outcomes.
The serving claim is architectural: the evaluated path makes zero request-time LLM calls and therefore avoids repeated online inference in the reranker, but total system cost is not measured.
The deployment controls in Table \ref{tab:deployment-controls} are launch guardrails rather than evidence that the method has already improved production KPIs.
The reported bootstrap intervals resample evaluated users within the same offline log slice; they do not account for exposure bias, logging-policy changes, catalog shifts, or counterfactual user response.

The empirical scope is limited to NCF and LightGCN candidate generators in one connected-TV VOD setting.
Generality across sequential, search, editorial, popularity, or ensemble candidate sources is future work.
The fixed score settings are evaluated configurations, not universal coefficients.
The main VOD dataset is proprietary and anonymized, which limits external reproduction.
We disclose the protocol, feature boundaries, metrics, scale bands, and aggregate results while withholding sensitive logs, identifiers, exact traffic volumes, exact profile-coverage rates, fallback proportions, and business-specific catalog details.

\section{Conclusion}

Ocean4Rec is a request-time-LLM-free reranking pattern for using offline LLM-derived OCEAN content profiles in VOD recommendation.
It materializes item OCEAN vectors from content metadata, aggregates recent implicit behavior into user OCEAN taste vectors, and reranks existing candidate lists with base relevance, OCEAN compatibility, and recency.
In same-candidate temporal-holdout offline evaluations on anonymized Samsung Smart TV VOD logs, Ocean4Rec has higher HR@20 and NDCG@20 point estimates than base-score-only sorting for both NCF and LightGCN candidate generators.
Compared with the stronger Base+Recency diagnostic, the added OCEAN signal improves NDCG@20 for both generators, is inconclusive for NCF HR@20, and is more pronounced for LightGCN.
The primary industry lesson is calibration: OCEAN is most useful as a bounded, inspectable auxiliary signal paired with recency and interaction-driven relevance.

\FloatBarrier

\bibliographystyle{ACM-Reference-Format}
\bibliography{Ocean4RecRecSysIndustry2026}

\end{document}